\documentclass[twocolumn,letter]{jpsj2}
\usepackage{graphicx}
\usepackage{amsmath,amssymb}

\def\runauthor{Youichi {\sc Yanase}}

\title{ 
FFLO Superconductivity near the Antiferromagnetic Quantum Critical Point
} 

\author{Youichi {\sc Yanase}\footnote{E-mail:
yanase@hosi.phys.s.u-tokyo.ac.jp}}

\inst{Department of Physics, University of Tokyo, Tokyo 113-0033, Japan
}

\recdate{Today 2008}

\abst
{ 
 The Fulde-Ferrel-Larkin-Ovchinnikov (FFLO) state near 
the antiferromagnetic quantum critical point (AFQCP) 
is investigated by analyzing the two dimensional Hubbard model 
on the basis of the fluctuation exchange (FLEX) approximation. 
 The phase diagram against the magnetic field and temperature is 
compared with that obtained in the BCS theory. 
 We discuss the influences of the antiferromagnetic spin fluctuation 
through the quasiparticle scattering, retardation effect, 
parity mixing and internal magnetic field. 
 It is shown that the FFLO state is stable in the vicinity of 
AFQCP even though the quasiparticle scattering due to the 
spin fluctuation is destructive to the FFLO state. 
 The large positive slope ${\rm d} H_{\rm FFLO}/{\rm d}T$ and the 
convex curvature (${\rm d}^{2} H_{\rm FFLO}/{\rm d}T^{2} > 0$) are obtained, 
where $H_{\rm FFLO}$ is the critical magnetic field for the second order 
phase transition from the uniform BCS state to the FFLO state. 
 These results are consistent with the experimental results in CeCoIn$_5$. 
 The possible magnetic transition in the FFLO state is examined. 
}

\kword
{
FFLO superconductivity; antiferromagnetic quantum critical point
}

\begin{document}
\sloppy
\maketitle

\newcommand{\eli}{$\acute{{\rm E}}$liashberg }
\renewcommand{\k}{\vec{k}}
\newcommand{\kk}{\vec{k'}}
\newcommand{\q}{\vec{q}}
\newcommand{\Q}{\vec{Q}}
\renewcommand{\r}{\vec{r}}
\newcommand{\e}{\varepsilon}
\newcommand{\ee}{\varepsilon^{'}}
\newcommand{\s}{{\mit{\it \Sigma}}}
\newcommand{\J}{\mbox{\boldmath$J$}}
\newcommand{\vv}{\mbox{\boldmath$v$}}
\newcommand{\Jh}{J_{{\rm H}}}
\newcommand{\LL}{\mbox{\boldmath$L$}}
\renewcommand{\SS}{\mbox{\boldmath$S$}}
\newcommand{\Tc}{$T_{\rm c}$ }
\newcommand{\Tcf}{$T_{\rm c}$}
\newcommand{\Hc}{$H_{\rm c2}^{\rm P}$ }
\newcommand{\Hcf}{$H_{\rm c2}^{\rm P}$}
\newcommand{\etal}{{\it et al.}: }
\newcommand{\PRL}{Phys. Rev. Lett. } 
\newcommand{\PRB}{Phys. Rev. B } 
\newcommand{\JPSJ}{J. Phys. Soc. Jpn. } 


 The BCS theory assumes the Cooper pairs which have zero total 
momentum.~\cite{rf:BCS} 
 The Cooper pairing with finite total momentum 
was predicted in 1960's by Fulde and Ferrel~\cite{rf:Fulde1964} and 
Larkin and Ovchinnikov.~\cite{rf:Larkin1964} 
 The FFLO state can be stabilized in the presence of 
the high magnetic field or the molecular field. 
 In addition to the U$(1)$-gauge symmetry the spatial symmetry 
is spontaneously broken in this novel quantum condensed state. 
 Although the FFLO state had not been experimentally observed 
in about 40 years after the theoretical 
prediction,~\cite{rf:Fulde1964,rf:Larkin1964} 
several evidences for the FFLO state 
have been obtained in the recent experiments.~\cite{rf:Matsudareview}  
 The FFLO superconductivity or superfluidity is attracting growing interests 
not only in the field of superconductivity~\cite{rf:Matsudareview} 
but also in the study of cold atom gases~\cite{rf:Zwierlein2006} 
and high density quark matters.~\cite{rf:Casalbuoni2004}

 Vast studies of the FFLO state were triggered by the discovery of 
novel superconducting (SC) phases in the strongly correlated electron 
systems.~\cite{rf:Matsudareview}  
 For example, a new SC phase has been identified 
in the heavy fermion superconductor CeCoIn$_5$
at low temperatures and high magnetic fields.~\cite{
rf:Radovan2003,rf:Bianchi2003,
rf:Watanabe2004,rf:Capan2004,rf:Kakuyanagi2005,rf:Martin2005,rf:Mitrovic2006,
rf:Kumagai2006,rf:Miclea2006,rf:Correa2007} 
 Experimental data for the FFLO state have been 
obtained in several organic materials, such as 
$\lambda$-(BETS)$_2$FeCl$_4$,~\cite{rf:Uji2006} 
$\kappa$-(BEDT-TTF)$_2$Cu(NCS)$_2$~\cite{rf:Singleton2000,rf:Lortz2007} 
and (TMTSF)$_2$ClO$_4$.~\cite{rf:Shinagawa2007,rf:Yonezawa2008} 
 In these materials the paramagnetic effect which is essential for the FFLO 
superconductivity is important because the orbital de-pairing 
effect~\cite{rf:Gruenberg1966} is suppressed by the heavy quasiparticle mass 
and/or the layered crystal structure.

 Another common characteristics in these compounds is 
the presence of the AFQCP in the vicinity of the SC phase. 
 Although the critical spin fluctuation significantly affects the electronic 
state near the AFQCP, the influence of AFQCP on the 
FFLO state has not been investigated from the theoretical 
point of view. 
 The main purpose of this paper is to examine the stability of 
the FFLO state near the AFQCP. 
 Although the FFLO state has been investigated 
by Yokoyama {\it et al.}~\cite{rf:Yokoyama2007} 
on the basis of the random phase approximation (RPA),  
the retardation effect, quasiparticle scattering and internal field, 
which play an essential role in our study, are neglected in the RPA.

 We analyze the two-dimensional Hubbard model in order to investigate 
the FFLO superconductivity near the AFQCP. 
 The model is described as follows, 
\begin{eqnarray}
  \label{eq:Hubbard-model}
  && \hspace{-10mm}
  H=\sum_{{\k},\sigma} \varepsilon(\k) 
  c_{{\k}\sigma}^{\dag}c_{{\k}\sigma}
  + U \sum_{i} n_{{i}\uparrow} n_{{i}\downarrow} 
  - h \sum_{i,\sigma} \sigma n_{{i}\sigma}. 
\end{eqnarray}
 We consider the square lattice and choose the following tight-binding 
dispersion, 
\begin{eqnarray}
  \label{eq:high-tc-dispersion}
  && \hspace{-10mm}
 \varepsilon(\k)=-2t(\cos k_{\rm x}+\cos k_{\rm y})
  +4t'\cos k_{\rm x} \cos k_{\rm y} -\mu, 
\end{eqnarray} 
in which the $d$-wave superconductivity is stabilized by the 
antiferromagnetic (AF) 
spin fluctuation.~\cite{rf:Scalapino1995,rf:Moriya2000,rf:Yanase2004} 
 In the following, we choose the unit of energy as $2t=1$ and fix 
$t'/t=0.25$. 
 The chemical potential is chosen so that the number density of electrons 
is $n=0.9$. 
 The last term in eq.~(1) describes the Zeeman splitting where 
$h=\frac{1}{2}g \mu_{\rm B} H$ with $H$ being the magnetic field. 
 As for the influence of the magnetic field, we here focus on 
the Zeeman splitting which is essential for the FFLO 
superconductivity.~\cite{rf:Fulde1964,rf:Larkin1964} 
 We assume a sufficiently large Maki parameter 
$\alpha = \sqrt{2} H_{\rm c2}^{\rm orb}/H_{\rm c2}^{\rm P}$ and ignore 
the orbital effect~\cite{rf:Gruenberg1966,rf:Tachiki1996,rf:Houzet2001,
rf:Adachi2003,rf:Ikeda2007,rf:Maniv2007,rf:Shimahara1997,
rf:Mizuhima-Ichioka2005} for simplicity.

 The model eq.~(1) is analyzed on the basis of the 
FLEX approximation~\cite{rf:FLEX} which is based on 
the Luttinger-Ward formalism.~\cite{rf:Luttinger1960} 
 The critical fluctuation near the AFQCP is taken into account in 
this approach.~\cite{rf:Moriya2000,rf:Yanase2004} 
 The thermodynamic properties, such as the free energy, condensation energy 
and optical integral have been investigated 
at zero magnetic field.~\cite{rf:Yanase2005}

 It is straightforward to extend the Luttinger-Ward formalism to 
the FFLO state. 
 We obtain the thermodynamic potential as follows, 
\begin{eqnarray} 
  \label{eq:thermodynamic-super} 
&&  \hspace{-10mm} 
  \Omega(T,\mu,h) = \Omega_{0}(T,\mu,h) + \Omega_{\rm F}
+ \Omega_{\rm B}, 
\\
  \label{eq:thermodynamic-superF} 
&&  \hspace{-10mm}
\Omega_{\rm F} = 
- \sum_{k} 
[\log\{{\rm det}\tilde{G}(k)^{-1}/{\rm det}\tilde{G}^{(0)}(k)^{-1}
\} 
\nonumber \\
&&  \hspace{-10mm}
+ \sum_{\sigma} G_{\sigma}(k) \Sigma^{\rm n}_{\sigma}(k) 
- F(k) \Delta^{\dag}(k) - F^{\dag}(k) \Delta(k)], 
\\
  \label{eq:thermodynamic-superB} 
&&  \hspace{-10mm}
\Omega_{\rm B} = \Phi[G_{\sigma},F,F^{\dag}]|_{\rm st}, 
\end{eqnarray} 
where $\Omega_{0}(T,\mu,h)= -T \sum_{\k,\sigma} 
\log[1+\exp\{-\beta(\e(\k) -\sigma h)\}]$ 
is the thermodynamic potential at $U=0$. 
 We describe the normal and anomalous Green functions as, 
\begin{eqnarray}
 && \hspace{-17mm}
 \tilde{G}(k)=
 \left(
 \begin{array}{cc}
   G_{\uparrow}(k_{+}) & F(k) \\
   F^{\dag}(k) & -G_{\downarrow}(-k_{-})
 \end{array}
 \right),  
\\
 &&  \hspace{-10mm}
 =
 \left(
 \begin{array}{cc}
   G^{\rm n}_{\uparrow}(k_{+})^{-1} & 
   \Delta(k) \\
   \Delta^{\dag}(k) & 
   -G^{\rm n}_{\downarrow}(-k_{-})^{-1} 
 \end{array}
 \right)^{-1}, 
\end{eqnarray}
where $G^{\rm n}_{\sigma}(k)^{-1}=
G^{(0)}_{\sigma}(k)^{-1} - \Sigma^{\rm n}_{\sigma}(k)$ with 
$G^{(0)}_{\sigma}(k)=({\rm i}\omega_n - \e(\k) + \sigma h)^{-1}$. 
 We have described $\vec{k}_{\pm}=\vec{k} \pm \q_{\rm F}/2$, 
and $\q_{\rm F}$ is the total momentum of Cooper pairs in the FFLO state. 
 For simplicity we assume the Fulde-Ferrel (FF) state where the SC order 
parameter has the spatial modulation  
$\Delta(\r)=\exp(\pm {\rm i} \q_{\rm F} \r)$. 
 Since we neglect the Larkin-Ovchinnikov (LO) state, the stability of 
the FFLO state is underestimated in the following results. 
 Note that the FF state has the same critical magnetic field \Hc as the 
LO state when the second order phase transition is assumed. 
 We will investigate the LO state in the vicinity of the AFQCP 
in another publication.~\cite{rf:Yanase2008}

 The normal and anomalous self-energies in eq.~(7) are obtained from 
the generating function $\Phi[G_{\sigma},F,F^{\dag}]$ as, 
\begin{eqnarray}
 \label{eq:derivative}
 && \hspace{-15mm} 
 \Sigma^{\rm n}_{\sigma}(k) = \frac{\delta \Phi}{\delta G_{\sigma}(k)}, 
\\
 \label{eq:derivative2}
 && \hspace{-15mm} 
 \Delta(k) = -\frac{\delta \Phi}{\delta F^{\dag}(k)}, 
\\
 \label{eq:derivative3}
 && \hspace{-15mm} 
 \Delta^{\dag}(k) = -\frac{\delta \Phi}{\delta F(k)}, 
\end{eqnarray}
by which the stationary conditions are satisfied 
as,~\cite{rf:Luttinger1960,rf:Yanase2005} 
\begin{eqnarray}
 \label{eq:variational-condition}
 &&\hspace{-10mm} 
\frac{\delta \Omega}{\delta \Sigma^{\rm n}_{\sigma}(k)}=
 \frac{\delta \Omega}{\delta \Delta(k)}=
 \frac{\delta \Omega}{\delta \Delta^{\dag}(k)}=0. 
\end{eqnarray}

 In the FLEX approximation the generating function is given as, 
\begin{eqnarray}
 \label{eq:Phi-FLEX}
&& \hspace{-12mm}
\Phi[G_{\sigma},F,F^{\dag}] = 
U n_{\uparrow} n_{\downarrow} + 
\sum_{q} 
\{
\log[1- U \chi_{\pm}^{0}(q)]
\nonumber \\ 
&& \hspace{-12mm}
+\frac{1}{2} \log[(1 - U \chi_{\rm F}^{0}(q))^{2} 
                 - U \chi_{\uparrow}^{0}(q) \chi_{\downarrow}^{0}(q)]
\nonumber \\
&& \hspace{-12mm}
+\frac{1}{2} U^{2} [\chi_{\rm F}^{0}(q)^{2} 
                   +\chi_{\uparrow}^{0}(q) \chi_{\downarrow}^{0}(q)]
+U [\chi_{\pm}^{0}(q) + \chi_{\rm F}^{0}(q)] \},  
\end{eqnarray}
 where the irreducible susceptibilities are described as, 
\begin{eqnarray}
  \label{eq:irreducible-susceptibility}
&& \hspace{-3mm}
\chi_{\pm}^{0}(q) = -\sum_{k} 
[G_{\uparrow}(k+q) G_{\downarrow}(k) + F(k+q) F^{\dag}(-k)], 
\nonumber \\
\\
&& \hspace{-3mm}
\chi_{\sigma}^{0}(q) = -\sum_{k} G_{\sigma}(k+q) G_{\sigma}(k), 
\\
&& \hspace{-3mm}
\chi_{\rm F}^{0}(q) = -\sum_{k} F^{\dag}(k+q) F(k).  
\end{eqnarray} 
 We have used the notations $\sum_{k}=T/N \sum_{\omega_n,\k}$ 
and $\sum_{q}=T/N \sum_{\Omega_n,\q}$ where 
$\omega_n=(2 n + 1) \pi T$, $\Omega_n=2 n \pi T$, 
$T$ is the temperature and $N$ is the number of sites. 
 The unit $\hbar=c=k_{{\rm B}}=1$ is used throughout this paper.

 We carry out the numerical calculation on the $N=64 \times 64$ sites 
and takes the Matsubara frequency $N_{\rm f}=2048$ for $T \geq 0.004$. 
 We keep the cutoff of Matsubara frequency as 
$(N_{\rm f} + 1) \pi T > 12.8$ for $T < 0.004$, 
which is larger than the band width $W=4$.

 In the following results, the dominant component of 
SC order parameter $\Delta(k)$ has the $d_{\rm x^{2}-y^{2}}$-wave symmetry. 
 The order parameter has an admixed odd frequency $d_{\rm x^{2}-y^{2}}$-wave 
component in the finite magnetic field 
owing to the violation of time reversal symmetry. 
 The even frequency $p$-wave component 
is also admixed with the $d_{\rm x^{2}-y^{2}}$-wave component 
in the FFLO state due to the breakdown of inversion symmetry 
like in the non-centrosymmetric 
superconductors.~\cite{rf:Edelstein1989,rf:Matsuo1994} 
 The extended $s$-wave component is also induced in the FFLO state 
unless $\q_{\rm F} \parallel$ [110]. 
 We will show that the $p$-wave order parameter plays 
an quantitatively important role for the stability of the FFLO state, 
while the other admixed components are negligible. 
 In the following we assume the FFLO modulation vector $\q_{\rm F}$ 
along the [110]-axis for the numerical accuracy, although a slightly 
larger condensation energy is obtained for $\q_{\rm F} \parallel [100]$. 
 We have confirmed that qualitatively the same results are obtained for 
$\q_{\rm F} \parallel$ [100]. 
 The direction of the modulation vector $\q_{\rm F}$ is actually 
affected by the orbital effect~\cite{rf:Adachi2003,rf:Ikeda2007} 
which is neglected in this paper.

\begin{figure}[htbp]
  \begin{center}
\includegraphics[width=7.5cm]{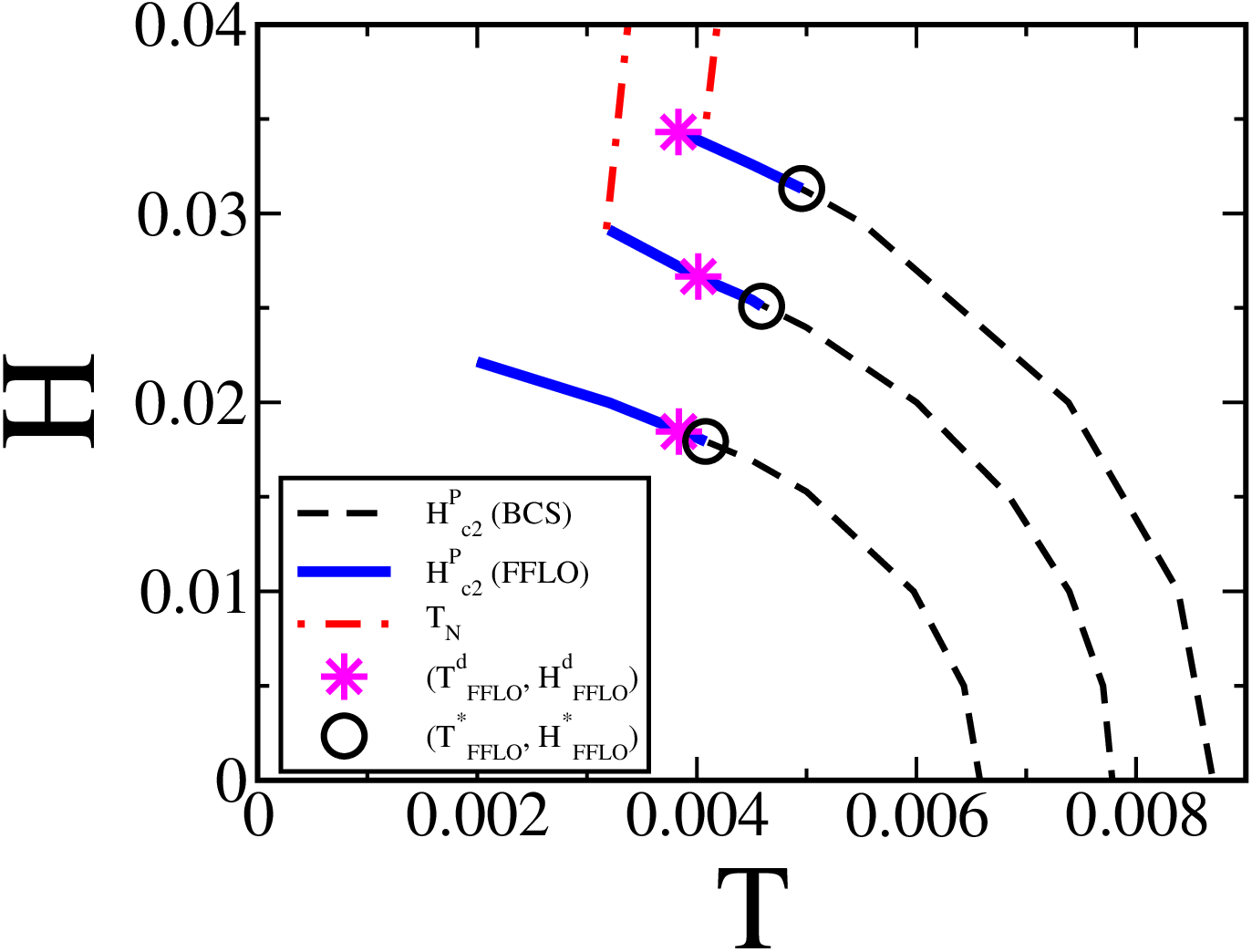}
\caption{(Color online) 
Paramagnetic critical magnetic field \Hc for $U/t=2.8$, $3.1$ and 
$3.4$ from the bottom to the top. The solid and dashed lines show 
the \Hc for the FFLO and BCS state, respectively. 
The tricritical point $(T_{\rm FFLO}^{*},H_{\rm FFLO}^{*})$ is shown 
by the circles ($\bigcirc$). 
 The stars show the tricritical point 
$(T_{\rm FFLO}^{\rm d},H_{\rm FFLO}^{\rm d})$ obtained by neglecting 
the $\q_{\rm F}$ dependence of the order parameter $\Delta(k)$. 
 The Neel temperature $T_{\rm N}$ is shown by the dash-dotted line 
for $U/t=3.1$ and $3.4$. 
}
    \label{fig:diagrams}
  \end{center}
\end{figure}

 First we investigate the tricritical point 
$(T_{\rm FFLO}^{*},H_{\rm FFLO}^{*})$ where the transition line 
from the FFLO state to the uniform BCS state merges to the 
critical magnetic field \Hcf$(T)$. 
 Figure~1 shows the results of \Hcf$(T)$ and 
$(T_{\rm FFLO}^{*},H_{\rm FFLO}^{*})$ for $U/t=2.8$, $3.1$ and $3.4$. 
 We show the transition temperature $T_{\rm N}$ from the normal state to 
the AF state in the same figure.~\cite{rf:Sakurazawa2005}  
 The Neel temperature $T_{\rm N}$ is determined by using the criterion 
$1- U \chi_{\pm}^{0}(q)=0.005$ at $\q = (\pi,\pi)$ and $\Omega_{\rm n}=0$, 
where the AF order occurs with the magnetic moment perpendicular to 
the magnetic field. 
 We find that the paramagnetic limit of the critical magnetic field 
\Hc is significantly enhanced by the internal field arising from the AF 
spin fluctuation. This is consistent with the experimental results in 
CeCoIn$_5$.~\cite{rf:Miclea2006}

 We see that the FFLO state is stable in the vicinity of the AFQCP.
 The temperature at the tricritical point $T_{\rm FFLO}^{*}$ in Fig.~1 
(see also Fig.~2) is larger than that obtained in the weak coupling 
BCS theory 
$T_{\rm FFLO}^{*}=0.56 T_{\rm c0}$~\cite{rf:Matsudareview,rf:Agterberg2001} 
where $T_{\rm c0}$ is the critical temperature at zero magnetic field. 
 This result seems to be surprising 
because the FFLO state is destabilized by the finite quasiparticle 
lifetime $\tau^{*}(\k)$~\cite{rf:Agterberg2001,rf:Adachi2003,
rf:Ikeda2007} which is significantly decreased by the 
critical spin fluctuation. 
 We here obtain the scattering rate 
$\Gamma_{\sigma}(\k)=1/\tau^{*}_{\sigma}(\k)=
-z_{\sigma}(\k){\rm Im}\Sigma^{\rm R}_{\sigma}(\k,0) 
\sim 2.3 T_{\rm c0}$ 
at $\k = (\pi,0)$ and $\sigma=\uparrow$ for $U/t=3.1$ and 
$(T,H)=(T_{\rm FFLO}^{*},H_{\rm FFLO}^{*})$. 
 Here, $z_{\sigma}(\k)=(1-
\partial {\rm Re} \Sigma^{\rm R}_{\sigma}(\k,\omega)/\partial \omega)^{-1}$ 
is the mass renormalization factor. 
 This value of $\Gamma(\k)$ is larger than the critical value 
$\Gamma \sim 1.4 T_{\rm c0}$ where the FFLO state disappears owing to the 
impurity scattering.

 The influence of the short lifetime is overcame by the other 
strong coupling effects, such as the retardation effect 
and the parity mixing. 
 We have confirmed that the frequency dependence of the SC order parameter 
$\Delta(k)$ significantly increases the ratio 
$T_{\rm FFLO}^{*}/T_{\rm c0}$. 
 The parity mixing between the $d_{\rm x^{2}-y^{2}}$-wave and $p$-wave 
order parameters furthermore enhances the FFLO superconductivity 
as shown in Fig.~1.  
 The stars in Fig.~1 show the tricritical point 
$(T_{\rm FFLO}^{\rm d},H_{\rm FFLO}^{\rm d})$ where 
the momentum and frequency dependences of the order parameter 
$\Delta(k)$ is fixed to be those in the uniform BCS state. 
 We see that the FFLO state is stabilized by the parity mixing, 
particularly near the AF critical point. 
 Thus, the FFLO superconductivity is stabilized by the strong coupling 
effects near the AFQCP. 
 This is consistent with the fact that the experimental indications 
for the FFLO superconductivity have been obtained in the strongly correlated 
electron systems near the AFQCP.~\cite{rf:Radovan2003,rf:Bianchi2003,
rf:Matsudareview,
rf:Uji2006,rf:Singleton2000,rf:Lortz2007,rf:Shinagawa2007,rf:Yonezawa2008}

\begin{figure}[htbp]
  \begin{center}
\includegraphics[width=5.5cm]{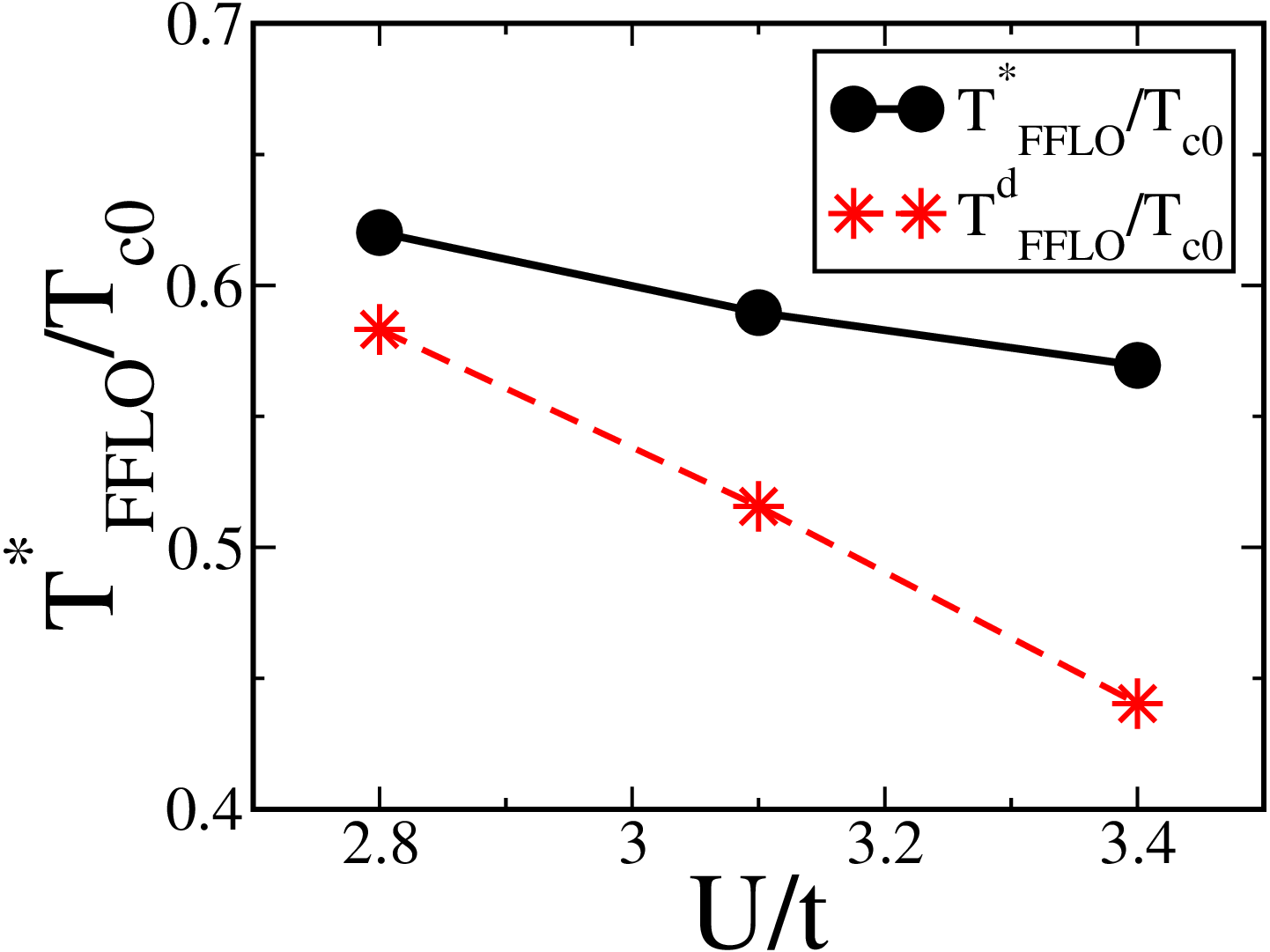}
\caption{(Color online) 
$U$-dependence of the ratio $T_{\rm FFLO}^{*}/T_{\rm c0}$ 
which represents the stability of the FFLO state. 
We show the $T_{\rm FFLO}^{\rm d}/T_{\rm c0}$ for a comparison. 
}
  \end{center}
\end{figure}

 Next, we discuss the $U$-dependence of the FFLO instability. 
 While the closeness to the AFQCP can be experimentally controlled 
by the pressure or carrier doping, that is represented by the 
parameter $U$ in our calculation. 
 Figure~2 shows the ratio $T_{\rm FFLO}^{*}/T_{\rm c0}$ which decreases 
as increasing $U$ and approaching to the AF instability. 
 This is mainly because of the quasiparticle scattering which is 
more sensitive to the closeness to the AFQCP than the other strong 
coupling effects. 
 Thus, the FFLO state becomes unstable little by little as approaching 
to the AFQCP. 
 This result is consistent with the pressure dependence of 
CeCoIn$_5$,~\cite{rf:Miclea2006} where $T_{\rm FFLO}^{*}/T_{\rm c0}$ 
gradually increases as increasing the pressure.

\begin{figure}[htbp]
  \begin{center}
\includegraphics[width=7cm]{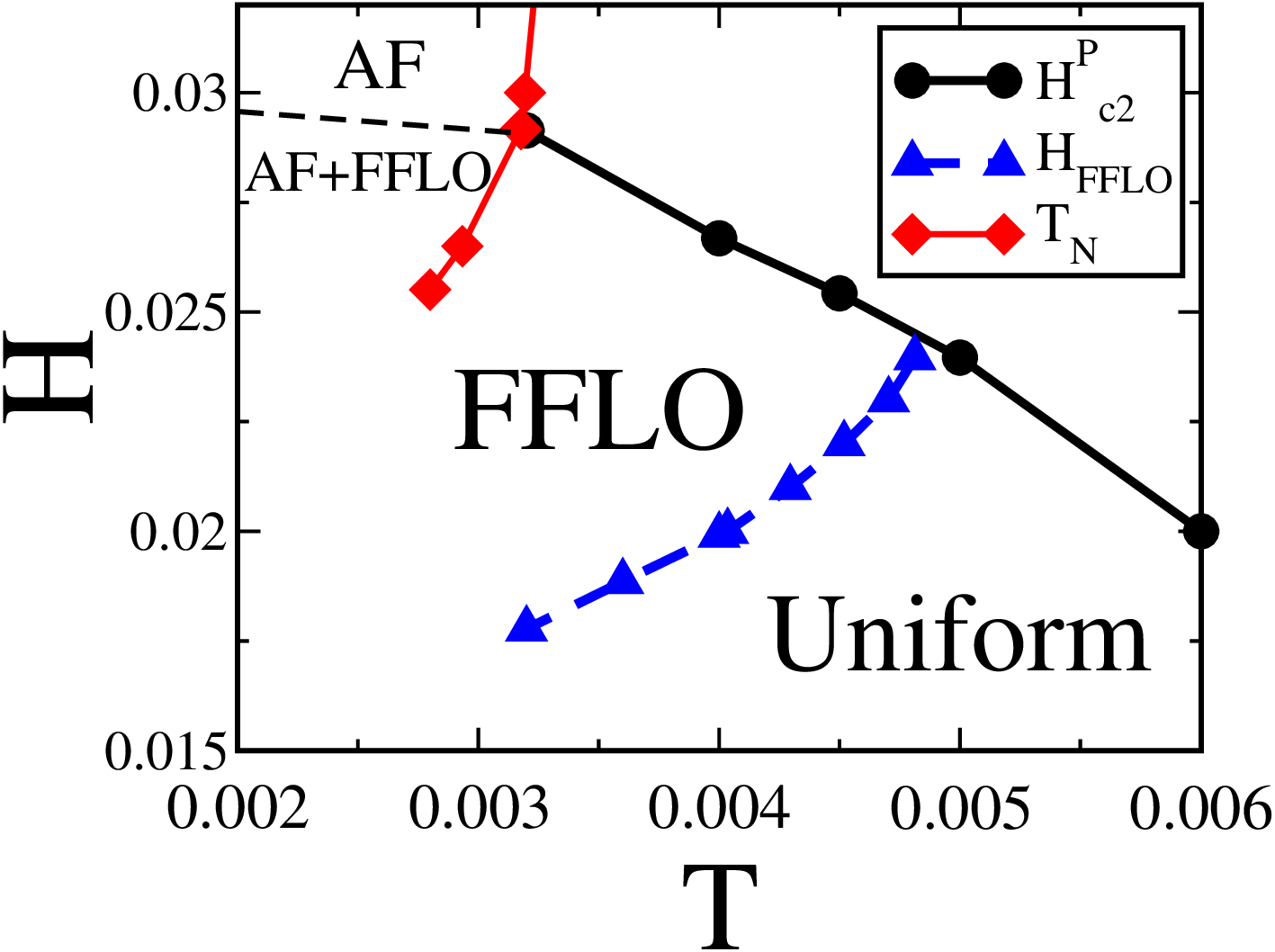}
\caption{(Color online) 
$H$-$T$ phase diagram for $U/t=3.1$ at high magnetic fields and 
low temperatures. 
Triangles show the second order phase transition 
from the uniform BCS state to the FFLO state. 
Diamonds show the second order phase transition to the AF state. 
We described the schematic thin dashed line below which the FFLO 
superconductivity coexists with the AF order. 
}
  \end{center}
\end{figure}

\begin{figure}[htbp]
  \begin{center}
\includegraphics[width=5.5cm]{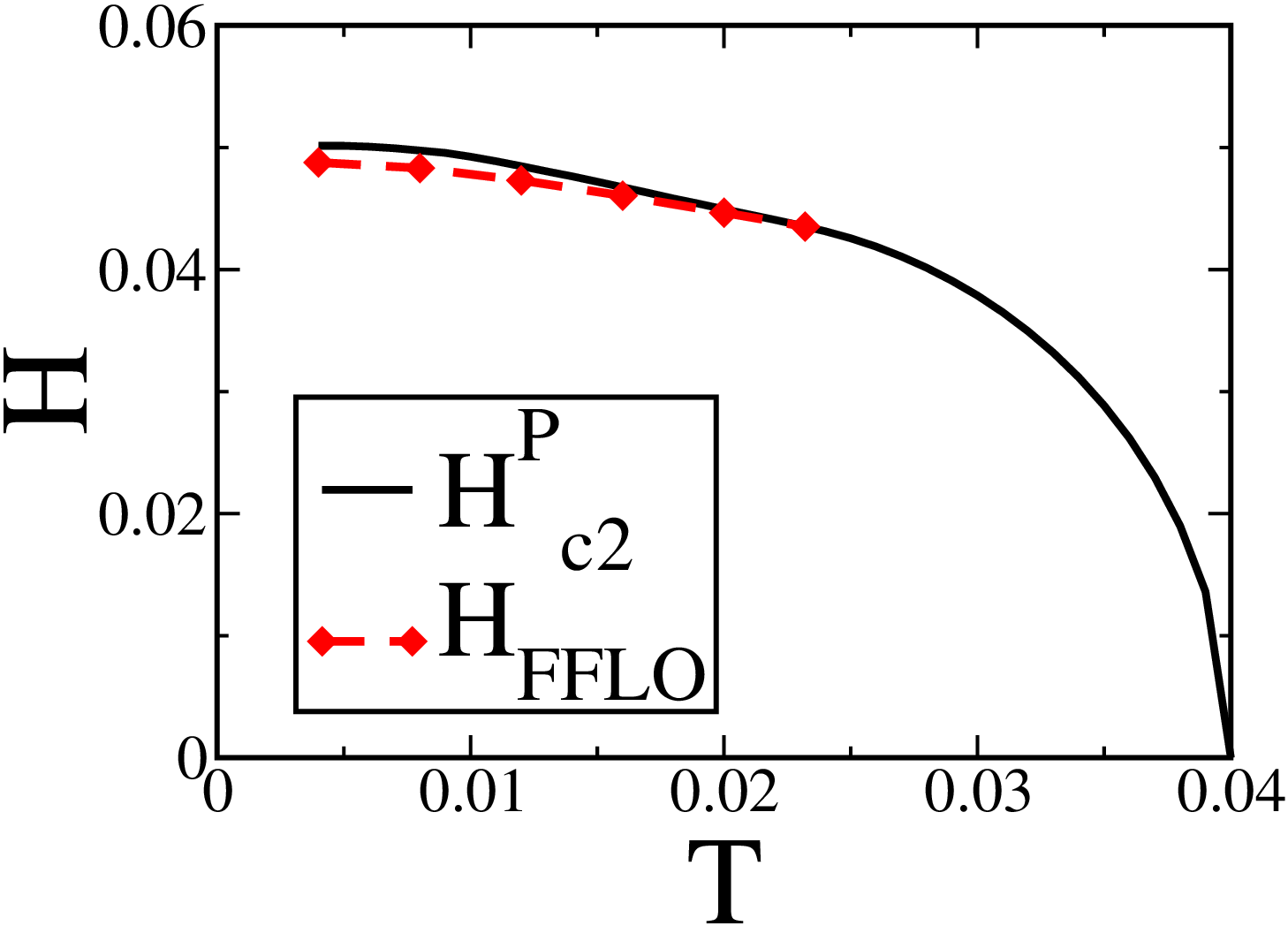}
\caption{(Color online) 
$H$-$T$ phase diagram of the BCS model eq.~(16). 
The solid line shows the \Hcf. The FFLO state is stabilized in the 
narrow region between the solid and dashed lines. 
}
  \end{center}
\end{figure}

 Finally, we investigate the phase transition from the uniform BCS 
state to the FFLO state. 
 Figure~3 shows the $H$-$T$ phase diagram for $U/t=3.1$. 
 It is shown that the critical magnetic field $H_{\rm FFLO}(T)$ 
from the uniform BCS state to the FFLO state rapidly decreases 
as decreasing the temperature. 
 The positive value of the slope ${\rm d} H_{\rm FFLO}/{\rm d}T$ 
is much larger than 
that in the weak coupling BCS theory.~\cite{rf:Vorontsov2005} 
 This is because the SC gap leads to the spin gap in the magnetic excitation 
and decreases the quasiparticle scattering rate. 
 FFLO superconductivity is enhanced below \Tc since the lifetime 
of quasiparticles significantly increases. 
 We obtain the convex curvature 
(${\rm d}^{2} H_{\rm FFLO}/{\rm d}T^{2} > 0$) which is consistent 
with the experiments in CeCoIn$_5$.~\cite{rf:Matsudareview,rf:Miclea2006} 
 The slope ${\rm d} H_{\rm FFLO}/{\rm d}T$ decreases as decreasing the 
temperature because the quasiparticle scattering rate almost vanishes 
at low temperatures in the SC state. 
 We note that the concave curvature 
${\rm d}^{2} H_{\rm FFLO}/{\rm d}T^{2} < 0$ is obtained in Ref.~31 
where the orbital effect is taken into account 
but the influences of the spin fluctuation are 
neglected.~\cite{rf:CommentAdachi} 
 Thus, the characteristics of the phase transition from the BCS state to the 
FFLO state in CeCoIn$_5$ are reproduced by taking into account 
the critical spin fluctuation, 
while they seem to be incompatible with the weak coupling 
BCS theories.~\cite{rf:Vorontsov2005,rf:Ikeda2007,
rf:Matsudareview,rf:Gruenberg1966}

 To illuminate the influence of the spin fluctuation more clearly
we show the H-T phase diagram in the BCS theory. 
 We obtain Fig.~4 by solving the following model, 
\begin{eqnarray}
  && \hspace{-10mm}
  H=\sum_{{\k},\sigma} \varepsilon(\k) 
  c_{{\k}\sigma}^{\dag}c_{{\k}\sigma}
  + g_{\rm s} \sum_{\k,\kk} B^{\dag}(\kk) B(\k) 
  - h \sum_{i,\sigma} \sigma n_{{i}\sigma}, 
\nonumber \\
\end{eqnarray}
on the basis of the BCS mean field theory. 
 Here, $B(\k) = \phi_{\rm d}(\k) 
c_{\k_{+},\uparrow} c_{-\k_{-},\downarrow}$ and 
$\phi_{\rm d}(\k)=\cos k_{\rm x} - \cos k_{\rm y}$. 
 This model has been used in the weak coupling theories for 
the $d$-wave FFLO state.~\cite{rf:Agterberg2001,rf:Ikeda2007} 
 We adopt the dispersion relation $\e(\k)$ as eq.~(2) and choose the 
attractive interaction $g_{\rm s} < 0$ so that $T_{\rm c0}=0.04$. 
 We assume the FF state and $\q_{\rm F} \parallel [110]$ to compare Fig.~4 
with Fig.~3 on an equal footing. 
 We see that the FFLO state is stable in the narrow region just 
below \Tc in Fig.~4. 
 Thus, the FFLO state is much more stable in the repulsive Hubbard 
model (Fig.~3) than the BCS model (Fig.~4). 
 This comparison illuminates the enhancement of the FFLO state near the 
AFQCP.

 The phase diagram in Fig.~3 seems to be consistent with the experimental 
results in CeCoIn$_5$, but there remains a discrepancy between our results 
and experiments. 
 Although the first order phase transition from the normal to SC state 
occurs in CeCoIn$_5$ above 
$T=T_{\rm FFLO}^{*}$,~\cite{rf:Bianchi2002,rf:Tayama2002} 
the SC transition is in the second order in our calculation for $T > 0.0032$. 
 This is because the quasiparticle scattering~\cite{rf:Agterberg2001} 
as well as the internal magnetic field~\cite{rf:Vorontsov2006} 
arising from the AF spin fluctuation 
suppress the first order phase transition. 
 Therefore, the first order phase transition in CeCoIn$_5$ above 
$T=T_{\rm FFLO}^{*}$ should be attributed to the other origin 
rather than the AFQCP. 
 For example, the orbital effect neglected in our calculation 
induces the first order phase transition above 
$T_{\rm FFLO}^{*}$.~\cite{rf:Adachi2003,rf:Ikeda2007} 
 Furthermore, we found that the unequal effective mass between two spin 
species in CeCoIn$_5$~\cite{rf:Onari2008,rf:Settai2001} 
also induces the first order phase transition, 
as will be shown in another publication.~\cite{rf:Yanase2008} 
 The heavy fermion physics, such as the strong electron correlation and 
the hybridyzation of localized and conduction electrons, may also affect 
the order of phase transition.

 At the last of this paper we comment on the magnetic instability 
in the FFLO state. 
 Figure.~3 shows the phase transition from the FFLO state to the AF state
(Diamonds) 
in contrast to the result at the zero magnetic field where 
the AF spin correlation decreases as decreasing the 
temperature below $T_{\rm c}$. 
 Our result in Fig.~3 shows that the AF order can coexist with the FFLO state 
due to the large residual density of states in the FFLO state. 
 We found that the AF order is furthermore enhanced in the LO state 
where the Andreev bound state at the spatial line node of 
SC order parameter induces the magnetic moment.~\cite{rf:Yanase2008} 
 These results may be consistent with the NMR measurement in 
CeCoIn$_5$,~\cite{rf:Young2007} although the discrepancy seems to remain in 
the experimental data.~\cite{rf:Young2007,rf:Kakuyanagi2005,rf:Mitrovic2006}

 In conclusion, we found that the FFLO superconductivity is stable 
near the AFQCP even though the quasiparticle scattering rate is 
significantly increased by the critical spin fluctuation. 
 The quasiparticle scattering is destructive to the FFLO state, but 
almost canceled by the strong coupling effect. The FFLO state is slightly 
suppressed as approaching to the AFQCP mainly because 
the quasiparticle scattering is increased. 
 The phase diagram obtained in our calculation is consistent with 
the experimental results in CeCoIn$_5$, 
however the first order phase transition is not reproduced. 
 We pointed out some possible mechanisms for the 
first order phase transition in CeCoIn$_5$. 
 The possible magnetic instability in the FFLO state was discussed.


 The authors are grateful to R. Ikeda, K. Izawa, Y. Matsuda 
for fruitful discussions.  
 Numerical computation in this work was carried out 
at the Yukawa Institute Computer Facility.


\begin{thebibliography}{9}
%

\bibitem{rf:BCS}
J. Bardeen, L. N. Cooper and J. R. Schrieffer: 
Phys. Rev. {\bf 108} (1957) 1175. 


\bibitem{rf:Fulde1964}
P. Fulde and R. A. Ferrel: Phys. Rev. {\bf 135} (1964) A550.


\bibitem{rf:Larkin1964}
A. I. Larkin and Yu. N. Ovchinnikov: Zh. Eksp. Teor. Fiz. {\bf 47} (1964) 
1136 [Sov. Phys. JETP {\bf 20} (1965) 762.] 


\bibitem{rf:Matsudareview}
Y. Matsuda and H. Shimahara: J. Phys. Soc. Jpn. {\bf 76} (2007) 051005 and 
references there in. 


\bibitem{rf:Zwierlein2006}
M. W. Zwierlein \etal 
Science {\bf 311} (2006) 492;
G. B. Partridge \etal
Science {\bf 311} (2006) 503. 

\bibitem{rf:Casalbuoni2004}
R. Casalbuoni and G. Nardulli: Rev. Mod. Phys. {\bf 76} (2004) 263.


\bibitem{rf:Radovan2003}
H. A. Radovan \etal
Nature {\bf 425} (2003) 51. 

\bibitem{rf:Bianchi2003}
A. Bianchi \etal
Phys. Rev. Lett. {\bf 91} (2003) 187004. 

\bibitem{rf:Watanabe2004}
T. Watanabe \etal
Phys. Rev. B {\bf 70} (2004) 020506(R). 

\bibitem{rf:Capan2004}
C. Capan \etal
Phys. Rev. B {\bf 70} (2004) 134513. 

\bibitem{rf:Kakuyanagi2005}
K. Kakuyanagi \etal
Phys. Rev. Lett. {\bf 94} (2005) 047602. 


\bibitem{rf:Martin2005}
C. Martin \etal
Phys. Rev. B {\bf 71} (2005) 020503.


\bibitem{rf:Mitrovic2006}
V. F. Mitrovic \etal
Phys. Rev. Lett. {\bf 97} (2006) 117002. 

\bibitem{rf:Kumagai2006}
K. Kumagai \etal
Phys. Rev. Lett. {\bf 97} (2006) 227002. 

\bibitem{rf:Miclea2006}
C. F. Miclea \etal
Phys. Rev. Lett. {\bf 96} (2006) 117001. 

\bibitem{rf:Correa2007}
V. F. Correa \etal
Phys. Rev. Lett. {\bf 98} (2007) 087001


\bibitem{rf:Uji2006}
S. Uji \etal
\PRL {\bf 97} (2006) 157001. 

\bibitem{rf:Singleton2000}
J. Singleton \etal
J. Phys. Condens. Matter {\bf 12} (2000) L641. 

\bibitem{rf:Lortz2007}
R. Lortz \etal
\PRL {\bf 99} (2007) 187002. 

\bibitem{rf:Shinagawa2007}
J. Shinagawa \etal
Phys. Rev. Lett. {\bf 98} (2007) 147002. 

\bibitem{rf:Yonezawa2008}
S. Yonezawa \etal
arXiv:0801.0484; arXiv:0804.1524. 


\bibitem{rf:Gruenberg1966}
L. W. Gruenberg and L. Gunther: \PRL {\bf 16} (1966) 996. 


\bibitem{rf:Yokoyama2007}
T. Yokoyama \etal
arXiv:0706.3270. 


\bibitem{rf:Scalapino1995}
D. J. Scalapino, Phys. Rep. {\bf 250} (1995) 329. 

\bibitem{rf:Moriya2000}
T. Moriya and K. Ueda, Adv. Phys. {\bf  49} (2000) 555.

\bibitem{rf:Yanase2004}
Y. Yanase \etal
Phys. Rep. {\bf 387} (2004) 1. 


\bibitem{rf:Tachiki1996}
M. Tachiki \etal
Z. Phys. B {\bf 100} (1996) 369. 

\bibitem{rf:Shimahara1997}
H. Shimahara and D. Rainer: \JPSJ {\bf 66} (1997) 3591. 


\bibitem{rf:Houzet2001}
M. Houzet and A. Buzdin: \PRB {\bf 63} (2001) 184521. 


\bibitem{rf:Adachi2003}
H. Adachi and R. Ikeda: \PRB {\bf 68} (2003) 184510. 


\bibitem{rf:Ikeda2007}
R. Ikeda: \PRB {\bf 76} (2007) 134504; 054517. 


\bibitem{rf:Maniv2007}
T. Maniv, V. Zhuravlev: arXiv:0712.3981v1. 



\bibitem{rf:Mizuhima-Ichioka2005}
T. Mizushima \etal
Phys. Rev. Lett. {\bf 95} (2005) 117003; 
M. Ichioka \etal 
Phys. Rev. B {\bf 76} (2007) 014503. 



\bibitem{rf:FLEX}
N. E. Bickers \etal 
Phys. Rev. Lett. {\bf 62} (1989) 961;


\bibitem{rf:Luttinger1960}
J. M. Luttinger and J. C. Ward, Phys. Rev. {\bf 118} (1960) 1417. 


\bibitem{rf:Yanase2005}
Y. Yanase and M. Ogata: \JPSJ {\bf 74} (2005) 1534. 


\bibitem{rf:Yanase2008}
Y. Yanase: unpublished. 


\bibitem{rf:Edelstein1989} 
V. M. Edelstein: Sov. Phys. JETP {\bf 68} (1989) 1244;
P. A. Frigeri \etal 
Phys. Rev. Lett {\bf 92} (2004) 097001; 
S. Fujimoto: J. Phys. Soc. Jpn. {\bf 76} (2007) 051008; 
Y. Yanase and M. Sigrist: J. Phys. Soc. Jpn. {\bf 76} (2007) 043712. 


\bibitem{rf:Matsuo1994} 
The parity mixing in the FFLO state has been pointed out in 
S. Matsuo \etal 
\JPSJ {\bf 63} (1994) 2499. 


\bibitem{rf:Sakurazawa2005}
The field induced AF order has been investigated in 
K. Sakurazawa \etal
\JPSJ {\bf 74} (2005) 271. 



\bibitem{rf:Agterberg2001}
D. F. Agterberg and K. Yang: 
J. Phys. Condens. Matter {\bf 13} (2001) 9259.


\bibitem{rf:Vorontsov2005}
A. B. Vorontsov \etal 
\PRB {\bf 72} (2005) 184501. 


\bibitem{rf:CommentAdachi}
 The large positive slope and the convex curvature of $H_{\rm FFLO}(T)$ 
have been obtained in Ref.~30, 
but the quartic correction which is important for $H_{\rm FFLO}(T)$ 
has been neglected. See Refs.~31 and 32. 



\bibitem{rf:Bianchi2002}
A. Bianchi \etal 
\PRL {\bf 89} (2002) 137002. 


\bibitem{rf:Tayama2002}
T. Tayama \etal 
\PRB {\bf 65} (2002) 180504. 


\bibitem{rf:Vorontsov2006}
A. B. Vorontsov and M. J. Graf: 
\PRB {\bf 74} (2006) 172504. 


\bibitem{rf:Settai2001}
R. Settai \etal
J. Phys.: Condens. Matter 13 (2001) L627;
A. McCollam \etal
Phys. Rev. Lett. 94 (2005) 186401. 


\bibitem{rf:Onari2008}
S. Onari \etal 
\JPSJ {\bf 77} (2008) 023703. 


\bibitem{rf:Young2007}
B.-L. Young \etal
\PRL {\bf 98} (2007) 036402. 

\end{thebibliography}
\end{document}